\documentstyle[preprint,aps]{revtex}
\begin{document}
\title{Sign reversal of the order parameter in $s$-wave superconductors}
\author{A.A.Golubov}
\address{Institute for Solid State Physics, 142432 Chernogolovka, Russia.}
\author{I.I. Mazin}
\address{Geophysical Laboratory, Carnegie Institution of Washington, 5251 Broad
Branch Rd., NW, Washington DC 20015, USA and Max-Planck-Institut\ f\"ur Festk%
\"orperforschung, Heisenbergstr.1, D-70569\ Stuttgart,\ FRG.}
\maketitle

\begin{abstract}
We show that in a superconductor where two or more bands cross the Fermi
level it is possible, in the framework of the conventional ($s$-wave) BCS
theory, that the sign of the superconducting gap is different on the
different sheets of the Fermi surface. At least one of two conditions has to
be satisfied: (1) The interband pairing interaction is weaker than the
Coulomb pseudopotential, while the intraband one is stronger, or (2) there
is strong interband scattering by magnetic impurities. In the case of YBa$_2$%
Cu$_3$O$_7$ we shall argue that the first condition is possibly satisfied,
and the second one very likely satisfied. In many aspects such a
sign-reversal $s-$wave superconductor is similar to a $d$-wave
superconductor, and thus demands revising recent experiments aimed to
distinguish between the $s-$ and $d-$wave superconductivity in this compound.
\end{abstract}

1. Recently, a number of experiments probing the relative phase of the order
parameter $\Delta $ on different parts of the Fermi surface in the
superconducting YBa$_2$Cu$_3$O$_7$  have been reported \cite
{illi,ibm,eth,jap,dynes,ibm2,Minnes,Wohlleben}. Some of them\cite
{ibm,dynes,Minnes} seem to indicate the conventional pairing state with $%
\Delta $ having the same sign over the whole Fermi surface, while others
suggest that $\Delta $ changes in sign, as consistent, e.g., with the $d$%
-pairing. The question of the symmetry of the superconducting state, and
thus of the interpretation of these experiments, is of crucial importance
for distinguishing between the conventional mechanism for superconductivity
and more exotic mechanisms, or among the unconventional theories themselves.

It is generally believed that the $s$-pairing is inconsistent with sign
reversal of $\Delta .$ This is not true. A simple counterexample is the case
of two concentric Fermi spheres, which have gaps (order parameters) of
opposite signs. Such a state has pure $s-$symmetry. The two-dimensional
analog are two coaxial Fermi cylinders. A Josephson contact between two such
superconductors, or with a conventional superconductor, may show an unusual
behavior, sometimes similar to the $d-$pairing case.

In this paper we shall show under which conditions the situation similar to
the examples above, which we shall call the Interband Sign Reversal of the
Gap (ISRG), can be  realized, and we will also argue that these conditions
are not at all exotic but are likely to be realized in YBa$_2$Cu$_3$O$_7$.
We shall also discuss briefly how ISRG can manifest itself in Josephson
tunneling, and we shall make a link to the existing experiments.

2. The extension of the BCS theory for two or more superconducting bands was
first worked out by Suhl, Matthias, and Walker\cite{MSW} and independently
by Moskalenko \cite{Mosc}, and later elaborated on by many. It was realized%
\cite{df-met} that the fact that several bands cross the Fermi level is not
sufficient to have considerable many-band effects in superconductivity. Only
when the bands in question have a very different physical origin, can a
substantial effect   appear.

This is the case in many high-$T_c$ cuprates. In particular, YBa$_2$Cu$_3$O$%
_7$ is known to have four sheets of the Fermi surface, all four having a
different physical origin\cite{Kanazawa}: One is formed by the chain $%
pd\sigma $ (seen by positron annihilation), another is an apical oxygen band
(seen in de Haas-van Alphen experiments), and the last two are bonding and
antibonding combinations of the two $pd\sigma $ plane bands (seen by
angular-resolved photoemission). Basing on the richness of the band
structure of YBa$_2$Cu$_3$O$_7,$ several groups pointed out that at least
the two-band\cite{kresin}, or probably the whole four-band\cite{eilat,genzel}%
, picture should be used to describe superconductivity in this system.
Various experiments have been interpreted as indicating two or more
different superconducting gaps.

We shall now remind the basic equations of the multiband BCS theory\cite
{MSW,df-met}: The Hamiltonian has the following form:%
$$
H=\sum_{i,k\sigma \alpha }\epsilon _{i,k}c_{i,k\alpha }^{*}c_{i,k\alpha
}-\sum_{ij,kk^{\prime }\alpha \beta }\frac{g_{ij}}2c_{i,k\alpha
}^{*}c_{j,k^{\prime }\alpha }c_{i,-k\beta }^{*}c_{j,-k^{\prime }\beta }
$$
where $\epsilon _{i,k}$ is the kinetic energy in the $i$-th band, $%
c_{i,k\sigma }^{*}$ and $c_{i,k\alpha }$ are corresponding creation and
annihilation operator, and $g_{ij}$ is the averaged pairing potential.

The order parameter $\Delta $ on the $i$-th sheet of the Fermi surface is
given by the equation
\begin{equation}
\label{gaps}\Delta _i=\sum_j\Lambda _{ij}\Delta _j\int_0^{\omega _D}dE\frac{%
\tanh (\sqrt{E^2+\Delta _j^2}/2k_BT)}{\sqrt{E^2+\Delta _j^2}},
\end{equation}
if the cut-off frequency $\omega _D$ is assumed to be the same for all
sheets. $T_c$ is defined in the usual way by the effective coupling
constant, $\log (2\gamma ^{*}\omega _D/\pi T_c)=1/\lambda _{eff},$ $\gamma
^{*}\simeq 1.78.$ The effective coupling constant $\lambda _{eff}$ in this
case is simply the maximal eigenvalue $\lambda _{\max }$ of the matrix $%
\Lambda _{ij}=g_{ij}N_j,$ where $N_j$ is the density of states at the Fermi
level (per spin) in the $j$-th band. $\Lambda _{ij}$ plays the role of the
coupling constant $\lambda $ in the one-band BCS theory. Note that
conventional (isotropic) $\lambda $ is also defined in terms of $\Lambda
_{ij}$: $\lambda =$ $\sum_{ij}\Lambda _{ij}N_i/N=\sum_i\lambda _iN_i/N$,
where the mass renormalization for the $i$-th band is $\lambda
_i=\sum_j\Lambda _{ij}$, and $N=\sum_iN_i$. Obviously $\lambda _{eff}\geq $ $%
\lambda ,$ which means thatdue to larger variational freedom $T_c$ in the
multiband theory is always larger than in the one-band theory. The two are
equal in isotropic case, i.e. when $g_{ij}$ does not depend on $i,j$. An
instructive example of the opposite case is the two-band model with $\Lambda
_{11}=\Lambda _{22}=\Lambda >0$, $\Lambda $$_{12}=\Lambda _{21}=-\Lambda .$
Then $\lambda =0$, while $\lambda _{eff}=2\Lambda $. Note that the last
value is the same as when $\Lambda $$_{12}=\Lambda _{21}=\Lambda $. The
physical reason is that although there is no solution of Eq.\ref{gaps} with $%
\Delta _1>0$, $\Delta _2>0$, there is an obvious solution with $\Delta
_1=-\Delta _2\neq 0.$ Near $T_c,$ the solution of Eq.\ref{gaps} is $\Delta
_2/\Delta _1=(\lambda _{eff}-\Lambda _{11})/\Lambda _{12},$ demonstrating
directly that the sign reversal of the order parameter, $\Delta _2/\Delta _1,
$ takes place\ {\it when nondiagonal matrix elements\ }$\Lambda _{12}${\it \
and\ }$\Lambda $$_{21}${\it \ are negative}. The fact that conventional BCS
theory (Eq.\ref{gaps}) allows for the ISRG solution, has never, to our
knowledge, been mentioned in the extensive literature existing on
multiband superconductivity\cite{but}.One can easily check that Eq.\ref{gaps}
may have a superconducting solution even for all $g_{ij}<0$, {\it i.e., }%
when no attractive interaction is present in the system. The condition for
that is $|g_{12}|>(|g_1|N_1^2+|g_2|N_2^2)/2N_1N_2$. This is similar to the
well-known fact that in a system with repulsion the superconductivity with
higher angular momenta ($p,$ $d$) is possible, because of the sign reversal
of the order parameter. The main difference is that in the example above the
symmetry of the superconducting state is the same as of the normal state.
Below we shall demonstrate that even a fully attractive interaction $%
g_{ij}\geq 0$ can lead to the sign reversal if (a) interband pairing
interaction is weaker than Coulomb preudopotential, (b) there is strong
interband scattering by magnetic impurities.

3. If $g$'s are electron-phonon pairing potentials, then Eq.\ref{gaps}
should be corrected for a Coulomb repulsion, which can be readily done\cite
{df-met} by substituting $g_{ij}\longrightarrow g_{ij}-U_{ij}^{*}\approx $ $%
g_{ij}-U^{*}$, where the effective Coulomb repulsion $U^{*}$ is
logarithmically renormalized in the same way as in one-band
superconductivity theory ($U^{*}$ is assumed to be independent on $i,j$). A
direct consequence of that is that if the interband electron-phonon coupling
is weak, the situation with a negative gap, $g_{ij}-U^{*}<0$, can easily be
realized because of the interband repulsion. We illustrate that by numerical
calculations presented in Fig.\ref{fig1}. In this calculations the following
parameters had been used: $g_{12}=g_{22}=0$, $N_1=4N_2$, and $g_{11}=N_1^{-1}
$ so that to have $\lambda =1$. This choice of parameters corresponds to the
plane and chain bands in YBa$_2$Cu$_3$O$_7,$ as discussed below in Section
5. Several facts draw attention: First, in this model $T_c$ decreases with
the increase of $\mu ^{*}=U^{*}N$ substantially slower than in a one-band
case when $T_c\rightarrow 0$ when $\mu ^{*}\rightarrow \lambda $. Second,
the order parameter induced in the second band (``chains'') is always
negative; Its absolute value reaches maximum when $|\Delta _1|=|\Delta _2|,$
{\it i.e}., at $U^{*}=g_{pp}N_p/2(N_p-N_c)$.

4. Following the standard way of including the impurity scattering in the
BCS theory\cite{AG}, one writes the equations for the renormalized frequency
$\tilde \omega _n$ and order parameter $\tilde \Delta _n$ ($n$ is the
Matsubara index), which completely define the superconductive properties of
the system (see, e.g. Ref. \cite{entel}):%
\begin{eqnarray}
\hbar \tilde \omega _{i,n}&=&\hbar \omega _n+\sum_{j,m}\frac{\hbar ^2\tilde
\omega _{j,m}}{2Q_{j,m}}(
\gamma _{ij}+\gamma _{ij}^s)\nonumber \\
\tilde \Delta _{i,n}&=&\Delta _i+\sum_{j,m}\frac{\hbar ^2\tilde \Delta
_{j,m}}
{2Q_{j,m}}(\gamma _{ij}-\gamma _{ij}^s)\nonumber \\
\Delta _i&=&\pi T\sum_{j,n}\Lambda _{ij}\tilde \Delta _{j,n}/{Q_{j,n}}.
\label{AG}
\end{eqnarray}
Here $\omega _n=(2n+1)\pi T$, $Q_{i,n}=\sqrt{\tilde \omega _{i,n}^2+\tilde
\Delta _{i,n}^2}$, $\gamma $$_{ij}$ is the scattering rate from band $i$
into band $j$ due to nonmagnetic impurities, and $\gamma $$_{ij}^s$ is the
same for magnetic impurities. Near $T_c$ Eqs.\ref{AG} can be solved
analytically \cite{unpub}. For two bands, in the linear in $\gamma ,\gamma _s
$ approximation, the solution redices again to Eqs. \ref{gaps}, with the
effective coupling matrix $\Lambda $:

$$
\tensor{\Lambda }_{eff}=\tensor{\Lambda }-\frac \pi {8T_{c0}}%
\tensor{\Lambda }\cdot \left(
\begin{array}{ll}
2\gamma _{11}^s+\gamma _{12}^s+\gamma _{12} & \gamma _{12}^s-\gamma _{12} \\
\gamma _{21}^s-\gamma _{21} & 2\gamma _{22}^s+\gamma +\gamma _{21}
\end{array}
\right) \cdot \tensor{\Lambda }.
$$
When all $\Lambda $'s are equal, the standard Abrikosov-Gorkov result is
recovered: $\delta \lambda \approx -\pi \lambda ^2(\gamma _{11}^s+\gamma
_{12}^s+\gamma _{21}^s+\gamma _{22}^s)/8T_{co}.$ The main point of the AG
theory\cite{AG} is that $\gamma $$^s$ enters equations for $\omega $ and $%
\Delta $ with opposite signs. That is why the magnetic impurities appear to
be pair-breakers, and the non-magnetic ones not. The above solution shows
that in the multiband case of Eqs.\ref{AG} this argument works only for the
intraband non-magnetic scattering ($\gamma _{ii}$ drop out), while all other
scattering rates are, in principle, pair-breaking{\it .}

An interesting special case is $\Lambda _{12},\Lambda _{21}\ll \Lambda
_{11},\Lambda _{22}.$ Then in the effective $\Lambda $ matrix nondiagonal
elements  $\Lambda _{ij}^{eff}(i\neq j)=\Lambda _{ij}+\pi \Lambda
_{ii}\Lambda _{jj}(\gamma _{ij}-\gamma _{ij}^s)/8T_{c0}$ can become
negative, if $\gamma _{ij}^s$ is sufficiently large. As discussed above,
this situation will lead to ISRG. In order to demonstrate this effect
quantitatively, we solved the Eqs.\ref{AG} in the Eliashberg approximation
numerically, using the following parameters: $\Lambda _{11}=1,\Lambda
_{22}=0.5,\Lambda _{12}=0.025,\Lambda _{21}=0.1$. This choice is again
inspired by the situation in YBa$_2$Cu$_3$O$_7$:The ratio of the densities
of states in the bonding and antibonding bands in YBa$_2$Cu$_3$O$_7$,
according to the band structure calculations, is about 2.5 \cite{eilat}, but
it is likely the calculations underestimate this value (see discussion
below). Correspondingly, we used $\gamma _{21}=4\gamma _{12,}\gamma
_{21}^s=4\gamma _{12}^s.$ The results for the low-temperature regime, $%
T<<T_c,$ are shown in Fig.\ref{fig2}. In accord with the condition derived
above, when the difference $\gamma _{12}^s-\gamma _{12}$ becomes larger than
some critical value (in this case, $0.042\pi T_c$), the second gap changes
sign. In other words, when attractive interband coupling is relatively weak
and the magnetic interband scattering is strong the system will choose to
have two gaps of the opposite signs, losing in pairing energy, but avoiding
the pair-breaking due to interband scattering.

5. Let us consider now two cases relevant for YBa$_2$Cu$_3$O$_{7-\delta },$
and their applications for the Josephson effect.

(1) It is believed by many (e.g., Ref.\cite{kresin}) that the chain
electrons would  not be superconducting, or only weakly superconducting, if
not for the ``proximity effect'' from the planes. In our language, this
means that $\lambda _c-\mu _c^{*}=(g_{cc}-U^{*})N_c\approx 0$, where $c$
stands for {\it chains}. Then the sign of the gap, induced in the chains,
will be determined by the sign of $g_{pc}-U^{*};$ in a quite likely case of $%
g_{cc}<U^{*},$ ISRG between the chain and the plane bands takes place.

(2) One can also look for the ISRG between the bonding ($b$) and antibonding
($a$) combinations of the two plane bands. According to the calculations\cite
{TB8} and experiment\cite{camp}, it is the {\it a} band which has Van Hove
singularities near the Fermi level. In the calculations the singularities
are bifurcated, which makes the density of states in the antibonding band
2.5 times larger than in the bonding band, and is exremely sensitive to the
warping of CuO$_2$ planes, thus resulting in strong electron-phonon
interaction. Experimentally, the singularities are even closer to the Fermi
level than in the calculations, and are extendended towards $\Gamma $-point.
If, as it is often claimed, this singularity plays a crucial role in
superconductivity, then the {\it a }band is the superconducting one, and the
superconductivity in the {\it b} band is induced. Consequently, one has the
situation similar to the above-described ``{\it p-c}'' scheme.

(3) Furthermore, the ISRG due to magnetic impurities may also be relevant
for YBa$_2$Cu$_3$O$_7.$ Let us assume that the main magnetic scatterers are
antiferromagnetic (AF) spin fluctuations on the plane Cu sites. For the
moment we assume these fluctuations to be static (see, however, the
discussion below). Inelastic neutron scattering studies\cite{rossat} show
that the AF correlations between the planes survive even in the fully
oxygenated samples, where the intraplanar correlations are virtually
non-existent (correlation length $\xi /a=0.84\pm 0.04).$ This is in direct
contradiction with the popular assessment that the intraplane AF
correlations are more important than those between the planes. To understand
the consequences of this fact we shall again consider the bonding and the
antibonding band. The former is even with respect to the $z\rightarrow -z$
reflection, and the latter is odd. The standard Hamiltonian for the magnetic
scattering is
$$
H_{ij}^s=-\sum_{{\bf R}}\sum_{\alpha \beta }<i\alpha |J({\bf r-R}){\bf S}_{%
{\bf R}}{\bf \sigma }|j\beta >,
$$
where {\bf S}$_{{\bf R}}$ is the spin of the impurity at point {\bf R}, and $%
\alpha ,\beta $ are spin indices. In case of two antiferromagntically
correlated impurities in the two planes, $\langle i\alpha |$ and $\langle
j\beta |$ must be of different parity to render non-zero $H_{ij}^s$. This
means that only $\tau _{ab}^s$ is non-zero.

In the previous paragraph we considered static impurities. In this context
``static'' means that the characteristic frequency of the AF fluctuations $%
\hbar \omega _{AF}\alt\pi T_c$ =25 meV. It is not clear yet how large $%
\omega _{AF}$ is in YBa$_2$Cu$_3$O$_{7-x}$. Detailed calculations will have
to include the proper frequency dependence of the magnetic susceptibility in
the same way as it is done with the phonons in the Eliashberg theory. It is
obvious, however, that the qualitative conclusions will not change.

6. Let us discuss the consequences for the Josephson effect separately for
each of the above cases. Rather than trying to explain the contradictory
experimental results reported so far \cite
{illi,ibm,eth,jap,dynes,ibm2,Minnes,Wohlleben}, we shall indicate some
qualitative predictions of our model.

The supercurrent density through a grain boundary in a two-band
superconductor can be written as $J_s=\sum_{ij}J_c^{ij}\sin \phi ^{ij},$%
where $J_c^{ij}$is the Josephson critical current density corresponding to
tunneling between the bands $i$and $j,$and $\phi ^{ij}$is the
gauge-invariant phase difference of order parameters $\Delta _i$ and $\Delta
_j$. In the simplest case of the Ginzburg-Landau regime \cite{ambar}

$$
J_c^{ij}=\pi \Delta _i\Delta _j/4eR_{ij}k_BT
$$
where $R_{ij}=(\hbar /e^2)(p_F/2\pi \hbar )^2/\langle D\rangle $ is the
tunneling resistance per unit area, $\langle D\rangle $ is the
angle-averaged transparency of the barrier and $p_F=min(p_{Fi},p_{Fj})$.
Following Geshkenbein and Larkin\cite{Gesh-Lar}, we obtain immediately that
if the order parameters have different signs, $sign(\Delta _i)=-sign(\Delta
_j)$, then in the stationary case ($J_s=0)$ the finite phase difference
appears, $\phi _{ij}=\pi $. This is similar to the ``$\pi -$contact''
considered by Bulaevskii et al. \cite{Bulaev}, but in our case it is due to
the sign reversal of the order parameter in different bands.

Generally, the total critical current, $J_c^{tot}$, depends on orientation
of the boundary relative to crystallographic axes because of the angular
dependence of $R_{ij}$. It can become negative for certain directions when
the contribution due to interband tunneling prevails. The condition is $%
J_c^{12}+J_c^{21}>J_c^{11}+J_c^{22}$ for the HTS/HTS junction and $%
J_c^{12}>J_c^{11}$ for the HTS/LTS one. To some extent this effect is
similar to that considered by Sigrist and Rice \cite{Sig-Rice} for the $d-$%
pairing, but some of our predictions differ qualitatively, as discussed
below.

a) Let us first consider the $c-p$ scenario. For a HTS/LTS junction the
tunnel resistance $R_{12}(\theta )$ depends strongly on the angle $\theta $
relative to the $b$ axis, namely $R_{12}(\theta )$ has a sharp minimum at $%
\theta =0$ due to the strong angle dependence of a barrier transparency $%
\langle D\rangle $ for a tunneling process. Moreover, according to band
structure calculations \cite{TB8} kinetic energy of carriers along the
chains is larger than that in plains, thus leading to larger $\langle
D\rangle $ values. As a result, $J_c^{tot}(\theta )$$<$$0$ for small $\theta
$, whereas for all other angles $J_c^{tot}(\theta )$$>$$0$. Therefore an
intrinsic $\pi $- phase shift will occur in this case between tunneling
along $a$-and along $b$-directions. Then a dc SQUID with junctions on $a$
and $b$ faces of a crystal, will show a $\Phi _0/2$ shift of a field
dependence $I_c(H)$. This effect was observed in Refs. \cite{illi,eth,jap}
and attributed to the $d_{x^2-y^2}$ pairing state. Another consequence is a
shift of a Fraunhofer pattern for a single junction formed on the corner of
a crystal, because $J_c$ changes sign along the junction , as was discussed
in Refs. \cite{illi,jap} for $d$ - pairing. Evidently, the same effect will
take place in the considered case.

We shall also mention that nonzero Josephson current, observed for $c$-axis
tunneling in {\it Pb}/insulator/{\it Y$_{1-x}$Pr$_x$Ba$_2$Cu$_3$O$_7$}
tunnel junctions in Ref.\cite{dynes} is contradictory to the $d_{x^2-y^2}$
symmetry. Indeed, nonzero Josephson current was predicted theoretically for
c-axis contact between s-wave and d-wave superconductors in Ref.\cite{Tanaka}%
, but only in the second order in boundary transparency $\langle D\rangle $
, therefore this model can not explain the rather large values of $I_cR_n$
products of the order of 1 meV observed in Ref.\cite{dynes}. On the other
hand, this observation is consistent with the suggested $c-p$ scenario. The
reason is that, contrary to the case of $d_{x^2-y^2}$ symmetry, an average
order parameter in the $ab$ plane is nonzero.

Interesting consequences appear for{\it \ HTS/HTS} (grain-boundary)
junctions. As follows from the above arguments, if $\theta =0$ in {\it only
one} of the grains, then the grain boundary is a $\pi $-contact, otherwise
it is a conventional one. Consider a closed contour crossing $N$ grain
boundaries. The flux quantization condition in zero external field reads $%
n\Phi _o=LI_s^{ij}+\sum_{m=1,N}(\Phi _o/2\pi )\phi _{ij}^{(m)}$, where $L$
is a self-inductance of a ring and $\phi _{ij}^{(m)}$is a phase difference
across $m$-th junction. Then it follows immediately, that if a contour
crosses an {\it odd number of }$\theta =0$ junctions, a spontaneous
magnetisation of a ring with half-integer flux quantum will occur, and when
it crosses an even number of $\theta =0$ junctions the flux quantum will be
integer.

Spontaneous magnetization with half-integer flux quantum in a three-junction
ring and with integer flux quantum in a two-junction ring was demonstrated
recently for YBa$_2$Cu$_3$O$_7$ in Ref\cite{ibm2}. In this experiment all
grain-boundaries were of $\theta =0$ type. Thus, the results \cite{ibm2} are
in agreement both with our proposal and with the $d$-wave scenario discussed
by Sigrist and Rice. To distinguish between these two explanations the
measurements for different grain orientations are necessary. At the same
time, the absence of angular dependence of $J_c$ observed in Ref.\cite{ibm}
for a number of different grain-boundary orientations in {\it YBCO} does not
contradict our scheme. Indeed, in \cite{ibm} all six grain boundaries have
had $\theta \neq 0$ which results in $J_c^{tot}(\theta )\simeq $$%
J_c^{11}(\theta )\simeq const.$

Another interesting phenomenon observed first in Bi-based HTS \cite{Khomskii}%
and more recently in {\it YBCO} \cite{Wohlleben} is the paramagnetic
Meissner effect (``Wohlleben effect''). The explanation was proposed in Refs.%
\cite{Sig-Rice,Khomskii,Khomskii2} in terms of intrinsic $\pi $-junctions
between weakly coupled superconducting grains, giving rise to spontaneous
orbital currents in arbitrary directions. An external magnetic field will
align those spontaneous current loops and can produce a net positive
magnetization. Therefore, existence of $\pi $-junctions between at least
some of superconducting grains is a key point for the Wohlleben effect. As
discussed above, such intrinsic $\pi -$junctions may exist in the considered
two band superconductor with interband gap sign reversal, thus leading to
the possibility of the Wohlleben effect.

b) For our second scenario a sign of the total critical current in an {\it %
HTS/LTS} and an {\it HTS/HTS} junction in any given direction depends
crucially on a relation between the current components $J_c^{12}$ and $%
J_c^{11}$, i.e. on corresponding tunnel resistances. From symmetry
considerations, there exists no fundamental reason for a sign change of $%
J_c^{12}-J_c^{11}$ vs $\theta $ in the {\it ab }plane. As a result, the
intrinsic phase shift between the two bands can not be detected by Josephson
experiments, similar to the discussed above {\it c-p} scenario. In all
junction geometries such an ISRG superconductor will behave like a
conventional {\it s}-wave one.

However, a possibility of the Wohlleben effect still exists: Each given
contour would include an even number of $\pi $-contacts, which in an ideal
case would compensate one another. But in reality the tunnel resistances of
these contacts, depending of the local state of each grain boundary, will be
different, so that the compensation would become incomplete.

We note, that this scenario may be relevant not only to {\it YBCO} but also
to all double-plane materials, like{\it \ Bi}-based compounds.

7. The main goals of the current paper were to demonstrate the possibility
of the existence of an $s$-wave superconductor with the sign-reversal of the
gap, to point out some factors which favor such a state in {\it YBa$_2$Cu$_3$%
O$_7$}, and to emphasize that some experiments interpreted as unambiguous
evidences for the $d$-wave pairing can, in fact, be explained by the
suggested ISRG $s$-pairing.

In addition to that, we would like to outline some implication of our
analysis to a model of spin-fluctuation induced superconductivity (Ref.\cite
{MP} and references therein). In this model, AF fluctuations are dynamic,
and serve as the intermediate bosons to give superconductivity. Only one
plane is considered, and, much in the same spirit as in our analysis, the
order parameters $\Delta ({\bf k})$and $\Delta ({\bf k+Q})$, where {\bf Q }%
is the AF vector ($\frac \pi a,\frac \pi b)$, are of the opposite signs,
which in the case of the {\it YBa$_2$Cu$_3$O$_7$} Fermi surface leads to the
$x^2-y^{2}$symmetry for $\Delta ({\bf k})$. Apparently, if one
considers two AF coupled planes, and two bands, {\it a }and {\it b}, then
only in the {\it a-b }channel does a non-zero pairing potential appear. This
is similar to the observation\cite{ole} that in {\it YBCO} the {\it gerade}
(with respect to $z\rightarrow -z$) phonons contribute to the intraband (%
{\it a-a} and {\it b-b}) coupling only, and the {\it ungerade }phonons to
the interband coupling only. Even without solving the corresponding
equations, one can immediately predict the results: Since in a bilayer there
is no problem having gaps of the same sign on a given sheet of the Fermi
surface, an $s$-wave solution must exist, with $\Delta _a$ and $\Delta _b$
having opposite signs. Direct numerical calculations show indeed that
Montoux-Pines model for a bilayer has a stronger instability in {\it s}%
-channel with ISRG than in {\it d}-channel \cite{licht}. Another interesting
point is that the intraband phonon pairing and the interband
spin-fluctuation pairing can coexist and even help each other.

\

{\bf Acknowledgements}{\it .} The authors are thankfull to O.V.Dolgov,
G.M.Eliashberg, D.I.Khomskii, A.I.Liechtenstein, E.G.Maksimov and M.R.Trunin
for many useful and stimulating discussions. A.A.G. acknowledges partial
support by Russian State Program HTSC under project N 93-194.

\begin{figure}\caption{Critical temperature and superconducting gaps in a
system with
induced superconductivity
in the second band, as a function of Coulmb
pseudopotential}\label{fig1}\end{figure}
\begin{figure}\caption{Superconducting gaps at $T=T_c/2$ in a two-band system
with interband scattering on magetic and non-magnetic impurities. Note the
straight
line corresponding to $\Delta_2=0$}\label{fig2}\end{figure}

\end{document}